\documentclass[aps,pre,amsmath,amssymb,lengthcheck,superscriptaddress]{revtex4-2}

\usepackage{graphicx}\graphicspath{ {figures/} }
\usepackage{hyperref}
\hypersetup{colorlinks,allcolors=blue,breaklinks}

\newcommand{\be}{\begin{equation}}
\newcommand{\ee}{\end{equation}}
\newcommand{\bs}{\begin{split}}
\newcommand{\es}{\end{split}}
\newcommand{\ba}{\begin{align}}
\newcommand{\ea}{\end{align}}
\newcommand{\bi}{\begin{itemize}}
\newcommand{\ei}{\end{itemize}}

\newcommand{\bla}{bla\\bla\\bla\\bla\\bla}

\begin{document}

\title{Minimal dissipation with viscoelastic baths in weakly driven processes}

\author{Pierre Naz\'e}
\email{pierre.naze@icen.ufpa.br}

\author{Fabrício Q. Potiguar}
\email{fqpotiguar@ufpa.br}

\affiliation{\it Universidade Federal do Par\'a, Faculdade de F\'isica, ICEN,
Av. Augusto Corr\^ea, 1, Guam\'a, 66075-110, Bel\'em, Par\'a, Brazil}

\date{\today}

\begin{abstract}

We investigate the thermodynamics of overdamped systems weakly driven by time-dependent protocols while interacting with viscoelastic heat baths. Using a generalized Langevin equation with memory, we derive the conditions under which the friction kernel ensures thermodynamic consistency, notably requiring the addition of a Dirac delta. Within linear response theory, we compute the relaxation function and relaxation time for two classes of protocols: a moving and a stiffening harmonic traps. Surprisingly, we find that viscoelastic memory does not always hinder relaxation; in certain cases, it accelerates it by reducing the effective relaxation time, leading to lower dissipation. We also derive optimal protocols that minimize the irreversible work and show how they are modified by the presence of the persistence time of the viscoelastic heat bath. Our results reveal that memory effects in the overdamped regime leave measurable thermodynamic signatures, depending on the protocol, with direct implications for controlling complex systems.

\end{abstract}

\maketitle

\section{Introduction}
\label{sec:introduction}

Many physical systems of interest, particularly in soft and biological matter, operate under conditions where inertia is negligible~\cite{van1992stochastic,risken1989fokker,seifert2025stochastic}. Examples include colloidal particles in aqueous environments~\cite{RusselSavilleSchowalter1989}, cellular components in the cytoplasm~\cite{SeksekBiwersiVerkman1997}, and tracer particles in complex fluids~\cite{MasonGanesanVanZantenWirtzKuo1997}. In such cases, the dynamics of the system are well-described by overdamped equations of motion, where friction dominates over acceleration. The overdamped limit not only simplifies the mathematical description but also captures the essential physics of slow, highly dissipative motion characteristic of mesoscopic systems.

In traditional Langevin theory, the friction is assumed to be local in time, corresponding to a Markovian bath with no memory. However, real-world environments often exhibit viscoelastic behavior, where the response of the medium depends on the past history of the system~\cite{Gomez-Solano_2014,Gomez-Solano_2015}. This memory is encoded in a nonlocal friction kernel, giving rise to a generalized Langevin equation~\cite{zwanzig2001nonequilibrium,mori65_Review}. While these equations are often studied in the underdamped regime, recent efforts have extended them to the overdamped limit, which is more appropriate for many experimental systems~\cite{pavliotis2021scaling}.

The incorporation of memory raises several theoretical challenges. Chief among them is the requirement of thermodynamic consistency: the modified dynamics must still obey the Second Law of Thermodynamics. This means that the memory kernel and the noise must satisfy a generalized fluctuation-dissipation theorem~\cite{gudowska2022fluctuation}. Furthermore, as we show in this work, ensuring positive dissipation in the overdamped limit often requires the addition of a Dirac delta term to the friction kernel. This term reflects an instantaneous dissipative response superimposed on the slower, history-dependent contribution.

In the context of stochastic thermodynamics, the focus is often on quantifying the irreversible work performed when a system is driven out of equilibrium by a time-dependent external parameter~\cite{naze2020compatibility,naze2022optimal,deffner2020thermodynamic}. In the linear response regime, this quantity is directly related to the system’s relaxation function, which describes how it returns to equilibrium after a perturbation. Thus, understanding how viscoelasticity alters relaxation is essential for predicting energetic costs and designing efficient control strategies.

While it is commonly expected that memory slows down the relaxation of a system—since the past resists the present change—we demonstrate that this intuition does not always hold in the overdamped limit. Under certain conditions, viscoelastic baths can actually accelerate relaxation, leading to a reduction in the total relaxation time. This effect, though subtle, has important thermodynamic consequences, as it implies a decrease in irreversible work for the same driving protocol.

In this work, we explore these ideas by analyzing two elementary yet insightful protocols: a moving harmonic trap and a stiffening trap~\cite{seifert2025stochastic}. Using the Ornstein–Uhlenbeck kernel as a representative model of viscoelasticity~\cite{uhlenbeck1930theory,caprini2024emergent}, we derive explicit expressions for the relaxation functions, relaxation times, and optimal driving protocols in each case. These examples allow us to isolate the effects of memory and show how they can either be suppressed or enhance the relaxation time, depending on the nature of the driving.

Our goal is to clarify how viscoelasticity manifests thermodynamically in the overdamped regime and to identify scenarios where its presence has measurable effects on dissipation and optimal control. In doing so, we contribute to a growing body of work~\cite{shrikant2023quantum,doerries2021correlation,loos2021stochastic} aimed at understanding non-Markovian thermodynamics and providing analytic tools for systems where memory cannot be ignored.

In Sec.~\ref{sec:wdp}, review the main concepts of linear response theory and introduce our notation. In Sec.~\ref{sec:2ndlaw}, we derive the conditions that a general friction kernel should have to satisfy the Second Law of Thermodynamics and to ensure a proper description of a physical system. Sec.~\ref{sec:examples} has the treatment of both examples cited. Sec.~\ref{sec:optimal}, we discuss the optimal work of both cases. Sec.~\ref{sec:conclusion} has our conclusions.

\section{Weakly driven processes}
\label{sec:wdp}

In this section, we will introduce the main concepts and notations used throughout this work.

Consider a system ruled by the generalized time-dependent overdamped Langevin equation~\cite{mori65_Review,zwanzig2001nonequilibrium}
\begin{equation}
 \gamma\int_0^t K(t-t')\dot{x}(t')dt'+\partial_x V(x(t),\lambda(t))=\eta(t),
 \label{eq:genlang}
\end{equation}
where $x(t)$ is the particle position, $\gamma$ is the damping coefficient, $V$ is the potential, $\lambda(t)=\lambda_0+g(t)\delta \lambda$ is the time-dependent external parameter, $g(t)$ is the protocol that controls the change of the parameter $\lambda(t)$, and follows $g(0)=0$ and $g(\tau)=1$, with $\tau$ the duration of the process, and $\eta(t)$ the viscoelastic noise, whose properties are
\begin{equation}
\label{eq:ounoise}
    \overline{\eta(t)}=0,\quad \overline{\eta(t)\eta(t')}=\frac{\gamma}{\beta} K(t-t'),
\end{equation}
where $\beta$ is the inverse effective temperature 
of the viscoelastic heat bath (VHB), $K(t)$ is the associated kernel. Observe that the system satisfies the second fluctuation-dissipation theorem and the heat bath is therefore passive~\cite{steffenoni2016interacting,maggi2017memory,shea2022passive,argun2016non,busiello2024unraveling}. The persistence time associated with the VHB is given by 
\begin{equation}
    \tau_P = \left.\mathcal{L}_s\left[\frac{K(t)}{K(0)}\right]\right|_{s=0},
\end{equation}
where $\mathcal{L}_s$ is the Laplace transform. The pure elastic heat bath (EHB) case (or Markovian/memoryless limit) is given by a kernel proportional to the Dirac delta, $K(t)=\delta(t)$, with persistence time $\tau_P=0$. Using linear response theory~\cite{kubo2012}, the relaxation function associated with the system is
\begin{equation}
    \Psi_0(t)=\beta\langle \overline{\partial_{\lambda_0}V(x(t),\lambda_0)} \partial_{\lambda_0}V(x_0,\lambda_0)\rangle_0-\mathcal{C},
\label{eq:relaxfunc}
\end{equation}
where the overline and the brackets indicate, respectively, averages over the noise and initial canonical distribution, given by
\begin{equation}
    \rho(x_0)\propto \exp(-\beta V(x_0,\lambda_0)).
\end{equation}
Here, $x_0=x(0)$ and $\lambda_0=\lambda(0)$. Also, $\mathcal{C}=\langle \partial_{\lambda_0}V(x_0,\lambda_0)\rangle_0^2$, which refers to the decorrelation of the relaxation function for long times~\cite{naze2020compatibility}. Here, $x(t)$ evolves accordingly to the solution of the non-perturbed system. In particular, the relaxation time associated with the relaxation function is~\cite{naze2020compatibility}
\begin{equation}
    \tau_R = \left.\mathcal{L}_s\left[\frac{\Psi_0(t)}{\Psi_0(0)}\right]\right|_{s=0}.
\label{eq:relaxtime}
\end{equation}

In linear response theory, the thermodynamic average work produced by a weakly driven process with $\delta\lambda/\lambda_0\ll 1$ is given by~\cite{naze2020compatibility}
\begin{equation}
    \langle W\rangle = \Delta F^{(2)}+\frac{1}{2}\int_0^\tau \int_0^\tau\Psi_0(t'-t'')\dot{\lambda}(t')\dot{\lambda}(t'')dt'dt'',
\end{equation}
where $\Delta F^{(2)}$ is the Helmholtz free energy difference between the initial and final states, expanded up to second order. In particular, we call
\begin{equation}
    W_{\rm irr} = \frac{1}{2}\int_0^\tau \int_0^\tau\Psi_0(t'-t'')\dot{\lambda}(t')\dot{\lambda}(t'')dt'dt''
    \label{eq:wirr}
\end{equation}
the irreversible work. Because of such a functional, the relaxation function must have its relaxation time and Fourier transform both non-negative to satisfy the Second Law of Thermodynamics~\cite{naze2020compatibility}, that is, $W_{\rm irr}\ge 0$. Also, significant progress has recently been made to find the optimal protocol $g^*(t)$ to minimize the thermodynamic average work and its fluctuation~\cite{schmiedl2007optimal, gomez2008optimal,naze2022optimal, naze2023optimal,naze2024analytical,davis2024active,deffner2020thermodynamic,blaber2023optimal,kamizaki2022performance}. In particular, in weakly driven processes, the optimal protocol is given by the following function~\cite{naze2024analytical}
\begin{equation}
\label{eq:optprot}
    g^*(t) = \frac{t+\tau_R}{\tau+2\tau_R}+\sum_{n=0}^\infty a_n\left[ \frac{\delta^{(n)}(t)-\delta^{(n)}(\tau-t)}{\tau+2\tau_R}\right],
\end{equation}
where $a_n$ is a coefficient related to the relaxation function given by
\be
a_{n} = -\left.\frac{d^n}{ds^n}\left[\frac{1}{s}\frac{1}{\mathcal{L}_s[\ddot{\Psi}_0(t)/\Psi_0(0)]}\right]\right|_{s=0}.
\ee
Also, the protocol is symmetric, with $g^*(t)=1-g^*(\tau-t)$. Observe that its form mainly depends on the ratio $\tau/\tau_R$ of the system, which will be a quantity highlighted in every example treated here. Indeed, this ratio indicates how close or distant the system is from equilibrium, that is, how much irreversible work is produced along the driving~\cite{naze2020compatibility}.

This work aims to characterize the relaxation function of systems ruled by Eq.~\eqref{eq:genlang}, when the VHB is presented. To do so, we find the conditions on the kernel to satisfy the Second Law of Thermodynamics. We study particular examples of a Brownian motion subject to moving laser and stiffening traps observing how they change under the influence of these new heat baths. Finally, using the previous results of Ref.~\cite{naze2024analytical}, we find the optimal protocol for this kind of system, showing that the presence of the VHB introduces Dirac deltas and derivatives in its form.

\section{Kernel compatibility with the 2nd Law}
\label{sec:2ndlaw}

In this section, we aim to find the properties of the friction kernel such that it becomes a useful function to model thermodynamic systems. This will be made considering its compatibility with the Second Law of Thermodynamics, that is, $W_{\rm irr}\ge 0$.

Consider the generalized overdamped Langevin equation~\eqref{eq:genlang} for the viscoelastic kernel $K(t)$ and generic potential. By integrating the generalized Langevin equation, the generic potential can be expressed as
\begin{equation}
\begin{split}
    V(x(t),\lambda(t))&=V(x_0,\lambda_0)+\int_0^t \eta(t')\dot{x}(t')dt'\\
    &-\gamma \int_0^t\int_0^{t'}K(t'-t'')\dot{x}(t')\dot{x}(t'')dt''
\end{split}
\label{eq:potential}
\end{equation}
Here, we make our first assumption by saying that the kernel is an even function. Therefore, we can write
\begin{equation}
\begin{split}
    V(x(t),\lambda(t))&=V(x_0,\lambda_0)+\int_0^t \eta(t')\dot{x}(t')dt'\\
    &-\frac{\gamma}{2} \int_0^t\int_0^{t}K(t'-t'')\dot{x}(t')\dot{x}(t'')dt''.
\end{split}
\label{eq:potential}
\end{equation}
We assume now that we are in linear response theory regime, considering only the non-perturbed solution of the above equation
\begin{equation}
\begin{split}
    V(x(t),\lambda_0)&=V(x_0,\lambda_0)+\int_0^t \eta(t')\dot{x}(t')dt'\\
    &-\frac{\gamma}{2} \int_0^t\int_0^{t}K(t'-t'')\dot{x}(t')\dot{x}(t'')dt''.
\end{split}
\label{eq:potential}
\end{equation}
Taking now the partial derivative in the initial parameter, multiplying then by $\partial_{\lambda_0} V(x_0,\lambda_0)$, taking the average on the viscoelastic noise and initial equilibrium distribution, and considering the necessary factors and constants, we arrive at the definition of the relaxation function
\begin{widetext}
\be
\Psi_0(t) -\Psi_0(0)= -\frac{\beta \gamma}{2} \int_0^t\int_0^t K(t'-t'')\langle \overline{\partial_{\lambda_0}(\dot{x}(t')\dot{x}(t''))}\partial_{\lambda_0}V(x_0,\lambda_0)\rangle_0dt'dt'',
\label{eq:kernel1}
\ee
\end{widetext}
where the term involving the correlation of the viscoelastic noise with $\partial_{\lambda_0}\dot{x}$ was considered null since this last term does not depend on the noise. The left-hand side of Eq.~\eqref{eq:kernel1} is always negative. Indeed, since $\Psi(t)$ is positive-definite, it is easy to see
\be
\int_0^t\int_0^t \ddot{\Psi}_0(u-v)dudv \le 0 \quad \Rightarrow \quad \Psi_0(t)\le \Psi_0(0).
\ee
Therefore, the quadratic form on the right-hand side must be positive. This is always possible if, and only if, the Fourier transform of the kernel is positive-definite~\cite{naze2020compatibility}. So, that is a condition for suitable kernels. It is expected as well that the kernel goes to zero for long times. According to Ref.~\cite{naze2020compatibility}, this happens if the persistence time is finite
\be
\tau_P <\infty.
\ee
That is another condition for a reliable kernel. Let us analyze some particular cases to explore the form of such a function. For $t\to 0^+$, Eq.~\eqref{eq:kernel1} is consistent. For $t\to\infty$, one can always rewrite
\begin{widetext}
\begin{align}
\Psi_0(0)&= \frac{\beta \gamma}{2} \int_0^\infty\int_0^\infty K(t'-t'')\langle \overline{\partial_{\lambda_0}(\dot{x}(t')\dot{x}(t''))}\partial_{\lambda_0}V(x_0,\lambda_0)\rangle_0dt'dt''\\
&= \frac{\beta \gamma}{2} \int_0^1\int_0^1 K(\infty(t'-t''))\langle \overline{\partial_{\lambda_0}(\dot{x}(t')\dot{x}(t''))}\partial_{\lambda_0}V(x_0,\lambda_0)\rangle_0dt'dt''\\
&= \frac{\beta \gamma}{2} \int_0^1\int_0^1 \delta(t'-t'')\langle \overline{\partial_{\lambda_0}(\dot{x}(t')\dot{x}(t''))}\partial_{\lambda_0}V(x_0,\lambda_0)\rangle_0dt'dt''\\
&= \frac{\beta \gamma}{2} \int_0^\infty\langle \overline{\partial_{\lambda_0}(\dot{x}^2(t'))}\partial_{\lambda_0}V(x_0,\lambda_0)\rangle_0dt'\\
&= \frac{\beta}{2} (\langle [\partial_{\lambda_0}V(x_0,\lambda_0)]^2\rangle_0-\langle \partial_{\lambda_0}V(x_0,\lambda_0)\rangle_0^2)),
\end{align}
\end{widetext}
where a change of variables was used to bring the infinite inside the kernel in the second step. The kernel goes to zero except if $t'=t''$, indicating a behavior of a Dirac delta. The factor two indicates something odd with the kernel. Indeed, such a function must be added with an extra Dirac delta to recover stationarity properties~\cite{nascimento2019non}. In this manner, one will have
\begin{widetext}
\begin{align}
\Psi_0(0)&= \frac{\beta \gamma}{2} \int_0^\infty\int_0^\infty [\delta(t'-t'')+K(t'-t'')]\langle \overline{\partial_{\lambda_0}(\dot{x}(t')\dot{x}(t''))}\partial_{\lambda_0}V(x_0,\lambda_0)\rangle_0dt'dt''\\
&= \frac{\beta \gamma}{2} \int_0^\infty\int_0^\infty [2\delta(t'-t'')]\langle \overline{\partial_{\lambda_0}(\dot{x}(t')\dot{x}(t''))}\partial_{\lambda_0}V(x_0,\lambda_0)\rangle_0dt'dt''\\
&= \beta \gamma \int_0^\infty\langle \overline{\partial_{\lambda_0}(\dot{x}^2(t'))}\partial_{\lambda_0}V(x_0,\lambda_0)\rangle_0dt'\\
&= \beta (\langle [\partial_{\lambda_0}V(x_0,\lambda_0)]^2\rangle_0-\langle \partial_{\lambda_0}V(x_0,\lambda_0)\rangle_0^2)).
\end{align}
\end{widetext}
Notice that, without adding the Dirac delta, we would have obtained half of the same result, which is inconsistent with the definition~\eqref{eq:relaxfunc} evaluated at $t=0$. Indeed, this extra term is a direct consequence of the compatibility with the Second Law of Thermodynamics. Finally, it is expected that the system achieves memorylessness when the persistence time goes to zero, $\tau_P\to 0^+$. Again, this is only possible by adding the extra Dirac delta such that the kernel, in the limit, tends to another Dirac delta, and that one is then doubled to match the factor two below the equation. Indeed
\begin{widetext}
\begin{align}
\lim_{\tau_P\to 0^+}\Psi_0(t)&= \Psi_0(0)+\lim_{\tau_P\to 0^+}\frac{\beta \gamma}{2} \int_0^t\int_0^t [\delta(t'-t'')+K(t'-t'')]\langle \overline{\partial_{\lambda_0}(\dot{x}(t')\dot{x}(t''))}\partial_{\lambda_0}V(x_0,\lambda_0)\rangle_0dt'dt''\\
&= \frac{\beta \gamma}{2} \int_0^t\int_0^t [2\delta(t'-t'')]\langle \overline{\partial_{\lambda_0}(\dot{x}(t')\dot{x}(t''))}\partial_{\lambda_0}V(x_0,\lambda_0)\rangle_0dt'dt''\\
&= \beta \gamma \int_0^t\langle \overline{\partial_{\lambda_0}(\dot{x}^2(t'))}\partial_{\lambda_0}V(x_0,\lambda_0)\rangle_0dt'\\
\end{align}
\end{widetext}
Observe that in such a limit, the solutions from which the system evolves change to the EHB case as well.

Therefore, the properties that guarantee a suitable kernel is: having the properties of being even, of having a positive Fourier transform, having a finite persistence time, having an added Dirac delta, and achieving memorylessness for $\tau_P\rightarrow 0^+$. An important viscoelastic kernel used in the literature that accomplishes these features is the Ornstein-Uhlenbeck one~\cite{uhlenbeck1930theory}, given by:
\be
K(t) = \frac{\exp{\left(-\frac{|t|}{\tau_P}\right)}}{\tau_P},
\ee
which will be the model we study in the next sections.

\section{Examples}
\label{sec:examples}

In what follows, we will analyze the relaxation function, the relaxation time, and optimal protocols of Brownian motions confined in laser traps. Our calculations were made considering the friction kernel
\be
K(t) \to \frac{\delta(t)+K(t)}{2} 
\ee
with the following fluctuation-dissipation theorem
\be
\overline{\eta(t)\eta(t')}=\frac{2 \gamma}{\beta} \left(\frac{\delta(t)+K(t)}{2} \right).
\ee
These are a requirement to recover the EHB case in the memoryless limit $\tau_P\to 0^+$.

\subsection{Brownian motion in moving trap}

Consider a Brownian motion confined in a moving trap, where the time-dependent potential is given by
\begin{equation}
    V(x(t),\lambda(t))=\frac{\omega_0^2}{2}{(x(t)-\lambda(t))}^2,
\end{equation}
where we adopt a unit mass and whose relaxation function for a EHB is~\cite{naze2022optimal}
\begin{equation}
    \Psi_0(t)=\omega_0^2\exp{\left(-\frac{|t|}{\tau_R}\right)},
\label{eq:psi0mov}
\end{equation}
with a relaxation time given by $\tau_R=\gamma/\omega_0^2$. Considering the Ornstein-Uhlenbeck kernel, and $\lambda_0=0$, the relaxation function of such a system will be 
\begin{equation}
    \Psi_0(t)=A_1\exp{\left(-\omega_1|t|\right)}+A_2\exp{\left(-\omega_2|t|\right)},
\end{equation}
with
\be
\omega_1 = \frac{1}{\tau_P}+\frac{1}{\tau_R}+\frac{\sqrt{\tau_R^2 +\tau_P^2}}{\tau_R\tau_P},
\ee
\be
\omega_1 = \frac{1}{\tau_P}+\frac{1}{\tau_R}-\frac{\sqrt{\tau_R^2 +\tau_P^2}}{\tau_R\tau_P},
\ee
\be
A_1 = \frac{\omega_0^2}{2}\left[1-\frac{\tau_R-\tau_P}{\sqrt{\tau_R^2+\tau_P^2}}\right],
\ee
\be
A_2 = \frac{\omega_0^2}{2}\left[1+\frac{\tau_R-\tau_P}{\sqrt{\tau_R^2+\tau_P^2}}\right]
\ee
which are all positive if all parameters are positive as well. Also, its Fourier transform is positive
\begin{equation}
    \hat{\Psi}_0(\omega)=\sqrt{\frac{2}{\pi}}\left(A_1 \frac{\omega_1}{\omega_1^2+\omega^2}+A_2\frac{\omega_2}{\omega_2^2+\omega^2}\right),
\end{equation}
meaning that the system is compatible with the Second Law of Thermodynamics. Finally, the system will have the following relaxation time
\be
\tau_R^B = \tau_R,
\ee
which does not depend on the persistence time of the VHB. For such a potential, the memory effects are not effectively accounted. Physically, this means that no memory is accumulated during the driving to act on a short-time dissipation produced by the Ornstein-Uhlenbeck kernel after the driving is ceased. This occurs because of the influence of the driving at the equilibrium position, which does not alter the dissipated term with the kernel. Indeed, due to its exclusive dependence on the velocity, a simple shift is completely ignored. The relaxation function of Eq.~\eqref{eq:psi0mov} is restored in the memoryless limit $\tau_P\rightarrow 0^+$.

\subsection{Brownian motion in stiffening trap}

\begin{figure}
    \centering
    \includegraphics[width=0.75\linewidth]{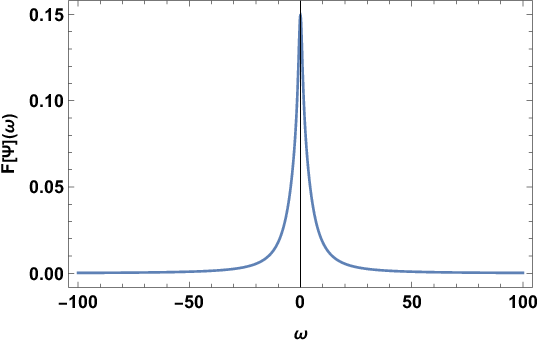}
    \caption{Fourier transform of the relaxation function of the stiffening trap in VHB. The function is positive, corroborating that the system is compatible with the Second Law of Thermodynamics. The unity was used for all parameters.}
    \label{fig:1}
\end{figure}

\begin{figure}
    \centering
    \includegraphics[width=0.75\linewidth]{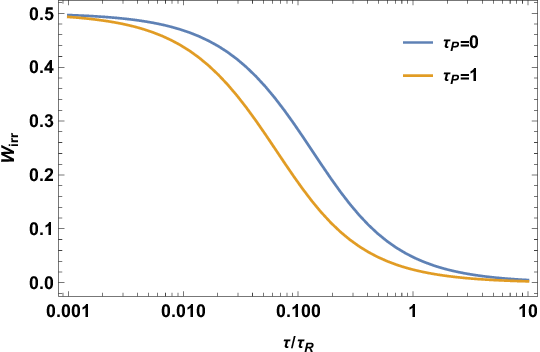}
    \caption{Comparison for the linear driving of the stiffening trap case for $\tau_P=0$ and $\tau_P=1$. The last one has less irreversible work than the first case, corroborating that the system is closer to the near-to-equilibrium case. It was used $\gamma=10$, $\omega_0=10$. The unity of the irreversible work is $\Psi_0(0)\delta\lambda^2$.}
    \label{fig:2}
\end{figure}

Consider now a Brownian particle confined in a stiffening trap, with the following time-dependent potential
\begin{equation}
    V(x(t),\lambda(t))= \lambda(t)\frac{x(t)^2}{2},
    \label{eq:stiffpotential}
\end{equation}
where we adopt a unit mass. In a purely elastic bath, the relaxation function of such a system is
\be
\Psi_0(t) = \frac{1}{2\beta \lambda_0^2}\exp{\left(-\frac{|t|}{\tau_R}\right)},
\label{eq:psi0stiff}
\ee
with $\tau_R=\gamma/(2\lambda_0)$~\cite{naze2022optimal}. Considering the Ornstein-Uhlenbeck kernel, the relaxation function of such a system will be
\begin{equation}
\Psi_0(t)=\exp{\left(-\omega_3|t|\right)}[A_3\cosh{\left(\omega_4|t|\right)}+A_4\sinh{\left(\omega_4|t|\right)}+A_5],
\end{equation}
with
\be
\omega_3 = \frac{2}{\tau_P}+\frac{1}{\tau_R},
\ee
\be
\omega_4 = \frac{\sqrt{4\tau_R^2 +\tau_P^2}}{\tau_R\tau_P},
\ee
\be
A_3 = \frac{1}{4\beta\lambda_0^2}\frac{8\tau_R^2-4\tau_R\tau_P+3\tau_P^2}{4\tau_R^2+\tau_P^2}
\ee
\be
A_4=\frac{1}{4\beta\lambda_0^2}\frac{4\tau_R-3\tau_P}{\sqrt{4\tau_R^2+\tau_P^2}}
\ee
\be
A_5=\frac{1}{4\beta\lambda_0^2}\frac{\tau_R\tau_P(12\tau_R+2\tau_P)}{(2\tau_R+\tau_P)(4\tau_R^2+\tau_P^2)}
\ee
Also, its Fourier transform is positive, meaning that it is composed of sums of positive parameters and $\omega^2$ (see Fig.~\ref{fig:1} for an example). Therefore, the system is compatible with the Second Law of Thermodynamics. Also, it has the following relaxation time
\be
\tau_R^B = \tau_R-\left[ \frac{20\tau_R^2+6\tau_P\tau_R}{(3\tau_R+4\tau_P)(4\tau_R+3\tau_P)}\right]\tau_P\ge 0,
\ee
meaning that the relaxation time is reduced compared to the EHB case. For this system, the decay is faster when associated with the Ornstein-Uhlenbeck kernel, which promotes a short-time dissipation in the system after the driving is ceased due to its exponential decay. Observe that the driving produces an effect in the velocity, contrasting with the moving laser trap case. Also, the reduction in the relaxation time produces an effect of reducing the dissipation of the system in the limit $\tau/\tau_R\gg 1$ indicates a system closer to the near-equilibrium case (see Fig.~\ref{fig:2}). The old relaxation function of Eq.~\eqref{eq:psi0stiff} is recovered in the memoryless limit $\tau_P\rightarrow 0^+$.

\section{Minimal work}
\label{sec:optimal}

\begin{figure*}
    \centering
    \includegraphics[width=0.3\linewidth]{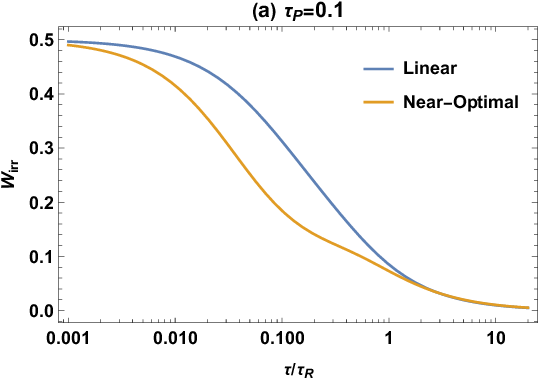}
    \includegraphics[width=0.3\linewidth]{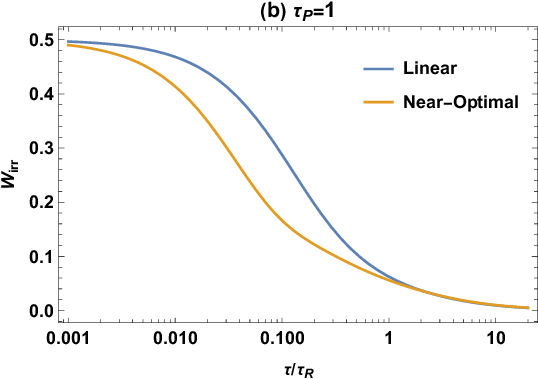}
    \includegraphics[width=0.3\linewidth]{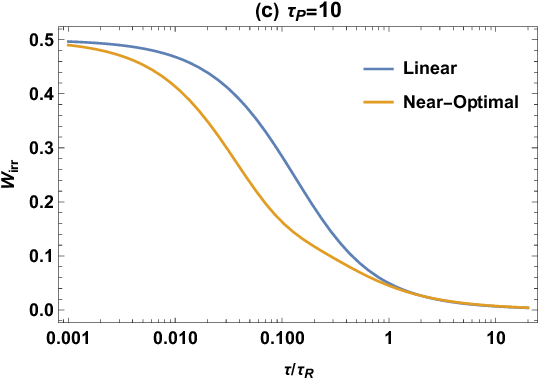}
    \includegraphics[width=0.3\linewidth]{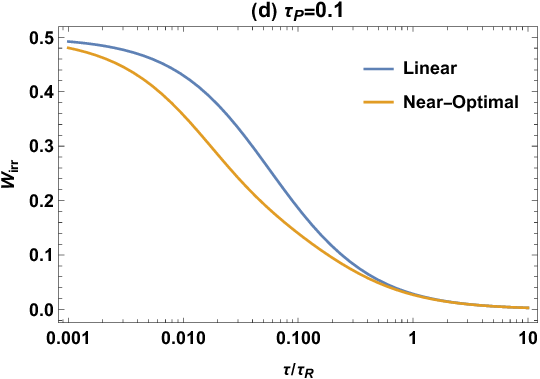}
    \includegraphics[width=0.3\linewidth]{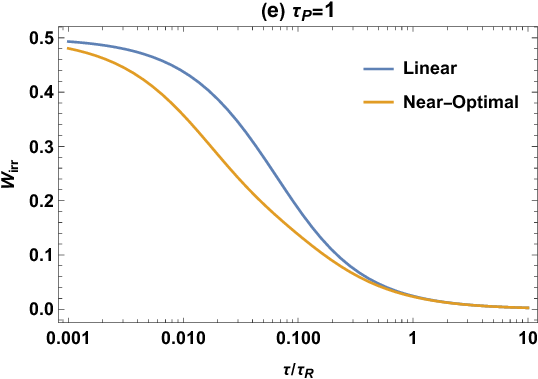}
    \includegraphics[width=0.3\linewidth]{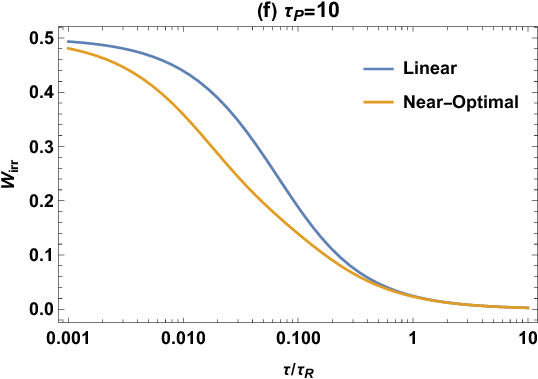}
    \caption{Minimal irreversible work for moving laser trap (panels (a), (b), (c)) and stiffening trap (panels (d), (e), (f)). In all cases, the continuous linear approximation of Eq.~\eqref{eq:nopt} produces a lesser irreversible work than the linear case of Eq.~\eqref{eq:linear}. For each case, the persistence time $\tau_P=0.1,1,10$ were used, corroborating no influence on the optimality. The values $\gamma=10$, $\omega_0=10$, $\beta=1$ were used. The unity of the irreversible work is $\Psi_0(0)\delta\lambda^2.$}
    \label{fig:3}
\end{figure*}

Regarding the optimal protocols for both examples, and upon inspection, it appears that all the terms of the Dirac deltas and their derivatives are necessary to describe their optimal protocols. Observing up to $n=3$, we observe all coefficients proportional to the persistence time $\tau_P$. Moreover, since we wait for a recovery of the EHB case in the memoryless limit, it is expected that all $a_n$ depend directly on monomials of $\tau_R$. Therefore, by contrast with the EHB case, which does not present delta terms, the VHB requires non-vanishing coefficients $a_n$, that eventually vanish in the EHB limit, $\tau_P\to 0^+$. In particular, the coefficient $a_0$ for the moving trap case is
\be
a_0 = -\frac{\tau_R\tau_P}{2},
\ee
and for the stiffening trap case
\be
a_0 = -4\tau_R\tau_P\left[\frac{112\tau_R^3+34\tau_R^2\tau_P+3\tau_R\tau_P^2}{(2\tau_R+\tau_P)(4\tau_R+3\tau_P)^2}\right].
\ee
The other terms can be seen on the repository. To corroborate the optimality of the result, in Fig.~\ref{fig:3}, we plot a comparison of the irreversible works between the linear protocol
\be
g(t)=\frac{t}{\tau},
\label{eq:linear}
\ee
and the near-optimal protocol
\be
g(t)=\frac{t+\tau_R^B}{\tau+2\tau_R^B}.
\label{eq:nopt}
\ee
Although it is just the linear part of the optimal protocol, it still achieves the desired reduction in the irreversible work. In particular, the limits for $\tau/\tau_R^B\to 0^+$ or $\tau/\tau_R^B\to \infty$ confirm the predicted result in Ref.~\cite{naze2024analytical}. Additionally, the optimality persists even with a change in the persistence time, corroborating that this parameter does not influence this aspect. In the calculation of the irreversible work, the jumps at the beginning and final point were considered~\cite{naze2022optimal}.

\section{Final remarks}
\label{sec:conclusion}

In this work, we have explored how viscoelastic baths influence the thermodynamics of weakly driven overdamped systems. By employing a generalized Langevin equation with memory and applying linear response theory, we derived consistency conditions that any admissible memory kernel must satisfy to remain compatible with the Second Law of Thermodynamics. A key requirement is the presence of an additional Dirac delta term in the friction kernel, which ensures positive dissipation and finite, well-defined relaxation behavior.

Our analytical treatment of two simple protocols—a moving trap and a stiffening harmonic trap—illustrates how viscoelastic effects impact relaxation and dissipation differently depending on the nature of the external driving. In the case of a moving trap, viscoelasticity does not alter the relaxation time, effectively masking memory effects. In contrast, for a stiffening trap, the bath contributes negatively to the relaxation time, leading to a faster decay of correlations and a reduction in irreversible work.

This finding is particularly interesting, as it reveals that memory does not always slow down system dynamics; under certain conditions, it can enhance relaxation. This acceleration is subtle and arises from the interplay between the instantaneous and delayed components of the kernel. It challenges the intuitive expectation that viscoelasticity necessarily introduces sluggishness and opens a new perspective on how structured baths may be harnessed to improve energetic efficiency.

Additionally, we derived analytical expressions for near-optimal protocols that minimize dissipation for a given driving time. These protocols explicitly depend on the relaxation function of the system, revealing how viscoelastic memory modifies the optimal temporal profile of control parameters. This demonstrates the importance of incorporating memory effects in the design of efficient driving strategies, especially in systems where precision and energetic efficiency are critical.

Overall, our results highlight that the thermodynamic signatures of memory are not only encoded in dissipation but also in the system’s ability to relax. These insights could prove useful in the design of optimal driving strategies in biological, colloidal, or engineered systems where viscoelastic environments are common. Future work may extend this framework to underdamped regimes and active matter, further enriching our understanding of non-equilibrium thermodynamics in complex media.

\section*{Data Availability}

The codes used to generate the data are available at \url{https://github.com/pnaze/OPAW}.

\bibliography{OPAW}

\end{document}